\documentclass{aa}
\usepackage{graphicx}

\newcommand{\msun}{$M_{\odot}\,$}
\begin{document}

\title{\bf RR Lyrae variables in Galactic globular clusters}
\subtitle{\bf V. The case of M3 pulsators. }

\author{ M. Castellani \inst{1} \and   V. Castellani
\inst{1}$^,$\inst{2} \and S. Cassisi \inst{3}}

\offprints {M. Castellani, \email {m.castellani@mporzio.astro.it}}

\institute{ INAF, Osservatorio Astronomico di Roma, via Frascati
33, 00040 Monte Porzio Catone, Italy \and INFN Sezione di
Ferrara,via Paradiso 12, 44100 Ferrara,Italy \and  INAF,
Osservatorio Astronomico di Teramo, Via Maggini, 00174 Teramo, Italy}

\date{Received ; accepted }

% THIS VERSION: RC5 - Release Candidate 5

%\authorrunning{Castellani et al.}

%\titlerunning{RR Lyrae stars in Galactic globular clusters. I.}

%\maketitle

%

%\markboth {Castellani et al.: RR Lyrae in globular clusters }

%{Castellani et al.:RR Lyrae in globular clusters }

\abstract{We use our synthetic Horizontal Branch (HB) procedure to
approach the debated problem concerning the adequacy of canonical
HB stellar models to account for the observed peaked distribution
of RR Lyrae fundamentalised periods in the globular cluster M3. We
find that by assuming a suitable bimodal mass distribution,
canonical models account for the observed period distribution. 
In particular, the best fit model, out of nine random extractions,
reaches a 99.9 \% Kolmogorov-Smirnov (KS)
probability.  We also
attempt a prediction of the relative distribution of variables in
fundamental and first overtone pulsators, reaching a rather
satisfactory agreement. However, one finds that canonical models
outnumber by roughly a factor of two the observed number of red HB
stars. Possible solutions for this discrepancy are outlined.
Alternative evolutionary scenarios are also briefly discussed.

\keywords {Stars: variables:RR Lyrae, Stars: evolution, Stars:
 horizontal-branch}
}

   \maketitle

%__________________________________________________

\section{Introduction}

In a previous paper of this series (Cassisi et al 2004: Paper
IV) we have presented the overall scenario of pulsational
predictions based on our synthetic Horizontal Branch (HB) procedure,
discussing the satisfactory  agreement between predictions
and selected properties of
RR Lyrae pulsators in globular clusters with various metallicities
and/or Horizontal Branch (HB) types. In this paper we will
approach a more detailed investigation, by discussing the case of
the RR Lyrae period frequency histogram in M3, which has been
recently claimed to be at variance with current predictions of stellar
evolution theory.

To shortly recall the history, Castellani \& Tornamb\'e (1981)
first draw the attention on the peaked distribution of fundamental
periods in M3 as an evidence requiring a peculiar distribution of
stars within the instability strip. The problem was revisited
by Rood \& Crocker (1989), who pointed out the difficulty to account
for the observed periods on the basis of smooth mass
distributions. More recently, the same problem has been addressed by
Catelan (2004), who reached the conclusion that the period
distribution of RR Lyrae variables in M3 is at odds with canonical
HB model predictions.

  The structure of the paper is the
following: in the next section we will discuss results based on our
synthetic procedure, showing  that
canonical HB models can closely reproduce the period frequency
histogram of RR Lyrae in M3 if a suitable bimodal distribution of
HB masses is assumed.
In Section 3 we will
attempt a prediction of the relative distribution of fundamental
and first overtone pulsators;
Section 4 deals with the distributions of stars
along the HB, discussing a troublesome disagreement between
observed and predicted number of red HB stars. A few comments
concerning alternative evolutionary scenarios will close the paper.

\section{M3: the models}

Concerning the evolutionary framework, we rely on the
stellar models already presented in Paper IV,
adopting for the cluster a metallicity Z=0.001, as in Catelan (2004).   
However,  
as outlined in Paper IV, one has to bear in mind 
that similar canonical models are
still affected by uncertainties, due to current limits in the
physical inputs. For example, changes in the adopted
$^{12}C(\alpha, \gamma)^{16}O$ nuclear cross section (Caughlan \&
Fowler 1985) affect both the morphology of the evolutionary track
and the core He-burning lifetime (Dorman 1992). In addition, all
stellar models whose evolutionary tracks are located in the cooler
portion of the HB, are affected by current uncertainty in the
efficiency of superadiabatic convection.

The  procedure adopted to compute synthetic RR Lyrae pulsators has
been exhaustively described in the previous papers of this series
and it will not be repeated here. Here we only note that present
work is based on up-to-date relations for both periods and
instability boundaries, as derived and discussed in Paper III (Di
Criscienzo, Marconi \& Caputo 2004). As a result, our  blue
instability boundary (BE), is located around $T_e \sim$ 7100-7200
K, i.e., about 200 K cooler than in Case A by Catelan(2004).
Moreover, the width of the instability strip is not a free
parameter but it is fixed by pulsational constraints at a
temperature cooler than the BE temperature by about $\Delta T_e
\sim 1200 K$.

However, the results presented in the following appear only
marginally dependent on the adopted pulsational scenario. As a
matter of fact, since the Van Albada \& Baker's (1971) formulation
of the relation connecting pulsation periods with the structural
parameters M, L and T$_e$, it was already clear that a smooth
distribution of stars within the instability strip will produce a
smooth distribution of periods, independently of any plausible assumption
about the instability boundaries and/or the star luminosity level
(see. e.g. Caputo \& Castellani 1975). As a consequence, it was
also clear that any smooth distribution of HB star masses, as
produced by the often adopted mass dispersion $\sigma_M \sim$ 0.02
\msun, was inadequate to account for the peaked period
distribution observed in M3.

%%%%%%%%%%%%%%%%%%%%%%%%%%%  FIGURA 1
\begin{figure}
\centering
\includegraphics[width=9cm]{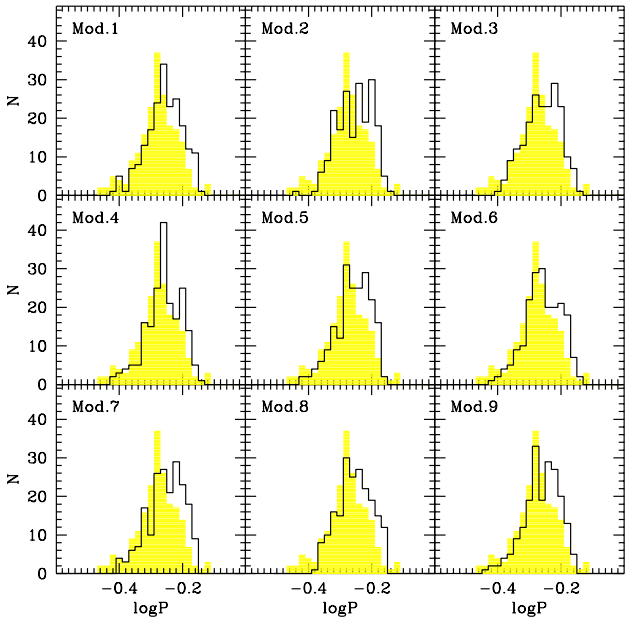}
%{Synt5_2bis.eps}
%\includegraphics[width=9cm]{m3f2nuova.eps}
\caption {Synthetic fundamentalised period frequency histograms
as obtained for a mean HB mass M=0.68 \msun with a dispersion
$\sigma_M \sim$ 0.005, and for nine different random extractions. The
shaded area shows the actual M3 distribution. } \label{f:fig1}
\end{figure}

%==========================TABLE 1=======================

\begin{table}
\caption{Values of KS probability for 9 runs, referred to different
samples: (a) RR Lyrae stars from the "red HB" only, (b) as (a) but
after the shift in luminosity, (c) adding the RR Lyrae from the "blue HB".
The last column gives the number of red HB. }

\begin{center}
\begin{tabular}{r r r r r }
\hline
$ Run $ & $ KS (a) $ & $ KS (b) $ & $ KS (c) $  & N(red)  \\
\hline

1 & 9.7 E-3 &98.55 & 92.24 & 246  \\
2 & 1.5 E-2 &13.24 & 22.62 &   302\\
3 & 1.7 E-1 &77.87 & 69.50 &  243\\
4 & 2.9 E-2 &91.54 & 99.93 & 240 \\
5 & 1.2 E-1 &44.86 & 41.93 &  219\\
6 & 1.8 E-1 &85.37 & 61.49 &  242\\
7 & 9.7 E-3 &20.78 & 21.52 &   213\\
8 & 8.2 E-2 &77.87 & 83.79 &    240\\
9 & 6.1 E-3 &69.59 & 60.78 &   201\\

\hline \hline
\end{tabular}
\end{center}
\end{table}
%==========================================================

To explore the predictions of canonical HB evolutionary tracks, we
decided - as a first step - to keep the assumption of a normal
deviate mass distribution, progressively decreasing the adopted
mass dispersion $\sigma_M$ and exploring the predicted period
frequency distribution with the mean mass as a free parameter. In
the case of a failure of such an approach, we were ready to test
different mass distributions. However, one finds that
the period distribution in M3 can be nicely reproduced when
assuming a mean HB mass M=0.68 \msun with a dispersion $\sigma_M
\sim$ 0.005.

Fig. \ref{f:fig1} shows the results of our first nine random
extractions as performed under the quoted assumption concerning
the mass distribution and by populating the HB till the number of RR
Lyrae observed in M3 was reached (see Castellani, Caputo \&
Castellani 2003: Paper I). The comparison with the observed
distribution, as given in the same figure, unambiguously
demonstrate that at least two out of the nine experiments (i.e.
simulations n.1 and 4) give a period frequency distribution which
closely resembles the observed one.

However, one easily recognizes that the predicted period
distribution appears marginally shifted toward larger values, by
an amount of the order of $\Delta logP \sim $0.02. Such a
shift is detected by the Kolmogorov-Smirnov test which gives a
quite small probability between observed and
predicted period distributions (Column 2 in Table 1).
According to the pulsation
theory (see Paper III) one can account for such a shift simply
decreasing the adopted HB luminosity by $\Delta logL \sim $0.02
or, alternatively by increasing the HB masses by $\Delta M \sim
0.04 M_{\odot}$.  We recall here that in Paper IV the HB
luminosity level was calibrated to reproduce the period interval
observed in M3. In this context, the above correction appears well
inside the uncertainty of that calibration.

\begin{figure}
\centering
\includegraphics[width=9cm]{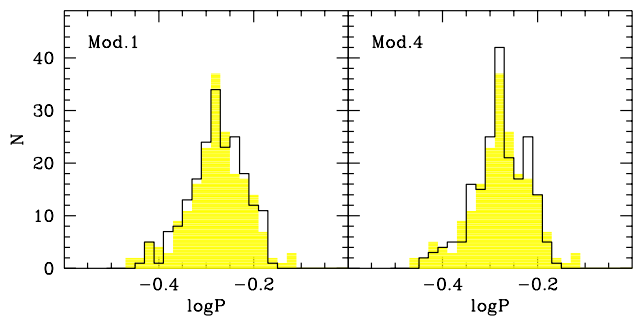}
\caption {The comparison between the observed period distribution
and the two best synthetic predictions (model n.1 and 4) 
as obtained from a gaussian distribution of the HB masses 
after the correction discussed in the text and before the addition of the blue component.}
\label{f:nuova}
\end{figure}

By applying such a correction one eventually finds that the
Kolmogorov-Smirnov test gives probabilities for the similarity of
observed and predicted distributions that sensitively increases,
ranging from 13.2\% (case n.2) to 98.6\% (case n.1). The column 3
in Table 1 gives the results of such a test, while Fig.
\ref{f:nuova} shows the comparison with the observed period
distribution and the two "best" predictions, as given by models
n.1 and 4. We conclude that canonical evolutionary tracks can
account for the peaked period histogram by adopting a suitable
mass distribution.

%%%%%%%%%%%%%%%%%%%%%%%%%%%  FIGURA

%%%%%%%%%%%%%%%%%%%%%%%%%%%

According to this simulation, RR Lyrae should be HB star
crossing the instability strip during their blueward  evolution
from the original red Zero Age Horizontal Branch (ZAHB) location. 
As a consequence,  the
simulation is producing  mainly red and RR Lyrae stars, with very
few stars hotter than the instability strip.  The evidence for a
rich population of blue HB stars in M3 thus requires the
additional contribution of a separate  population of less massive
HB stars, located at the blue side of the instability strip.

One may easily predict that such an additional population should
give a marginal contribution to the bulk of RR Lyrae pulsators,
since only stars in the later phases of HB evolution will cross
the instability strip. However, we will follow the suggestion of
our referee discussing this point in more details. As already done
by Catelan, we took from stellar counts the number of blue HB
stars, populating the blue HB portion with 206 stars uniformly
distributed over the range of masses 0.65 to 0.61 M$_{\odot}$. 
This means that we account for a HB population whose ZAHB temperatures
range from $logT_e$=3.859 to $logT_e$=4.045, i.e., from a
temperature close to the BE up to the temperature of the hot end
of the bulk of M3 blue HB stars. Note that the adopted
distribution maximizes the contribution of the blue HB population
to the variables. The same number of blue stars with, e.g., a
gaussian distribution around the mean mass 0.63 M$_{\odot}$ and a
dispersion  $\sigma_M \sim$ 0.01, covers a similar range of
temperatures  giving a lower number of pulsators, since a large fraction
of HB stars located close to the blue instability boundary moved towards hotter
colors.

\begin{figure}
\centering
\includegraphics[width=9cm]{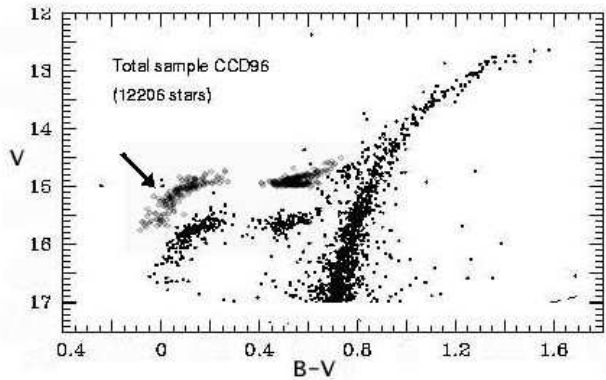}
\caption {The V, B-V color-magnitude diagram
of M3 (Ferraro et al. 1997) compared with
the predicted distribution of HB stars (arrow)
from the best model (model 4), as arbitrarily
shifted in magnitudes.}
\label{f:fig2}
\end{figure}

Using once again a set of 9 different random extractions we found
that the number of RR Lyrae from this  population  ranges from 7
to 20, i.e., giving a marginal but not negligible contribution to
the pulsator population. Figure \ref{f:fig2} shows the comparison
between the predicted color magnitude diagram distribution and
observational data from Ferraro et al. (1997). The simulations
account for an observational gaussian error with $\sigma$=0.02 in
both magnitudes. Column 4 in Table 1 shows that the inclusion of
such a contribution raises in the best case the KS probability up
to 99.9\%, supporting the occurrence in M3 of a
bimodal distribution of HB masses.

%%%%%%%%%%%%%%%%%%%%%%%%%%%  FIGURA 2 nuova

%%%%%%%%%%%%%%%%%%%%%%%%%%%

Such a bimodal distribution is of course an unexpected feature.
However, we notice that within the galactic globular cluster
family this feature is far from being unusual, since Harris (1974)
brought forward the striking bimodal distribution of HB stars in
the galactic globular NGC2808 (see, e.g., Catelan 2004; D'Antona
\& Caloi 2004, and references therein). According to the result of
the present investigation, one should conclude that even in the M3
case we are facing with a bimodal HB, though with a hidden
bimodality which becomes evident only when the 
distribution of RR Lyrae periods is taken properly into account.

\section{Fundamental and first overtone pulsators}

By relying on the overall agreement between observed and predicted
fundamentalised periods, one can go deeper in the comparison, thus
testing the predictions of the theoretical scenario concerning the
relative distribution of fundamental and first overtone pulsators.
To this purpose we adopted  from pulsational theories the topology
of the instability strip, with first overtone pulsators in the
hotter portion of the strip, fundamental pulsators in the cooler
portion, and an intermediate range of temperatures where both
modes can be stable (the OR zone).

\begin{figure}
\centering
\includegraphics[width=9cm]{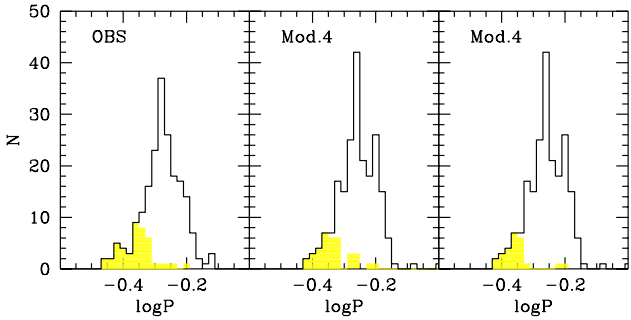}
\caption {The observed distribution of  fundamentalised periods in
M3 (left panel) as compared with the best synthetic theoretical
distributions obtained assuming an hysteresis mechanism (middle
panel) or a fixed transition temperature (right panel). The shaded
areas  show the contribution of first overtone (RRc) pulsators.}
\label{f:fig3}
\end{figure}

Fig. \ref{f:fig3} shows that the synthetic models, using the
recipe of the hysteresis mechanism (Van Albada \& Baker 1971),
i.e., by assuming that in the OR zone the variables pulsate in
their previous pulsation mode, can reach a reasonable similarity
with observations, provided that theoretical estimates for the
fundamental blue boundary are decreased by about 100 K, i.e., down
to $logT_e$=3.832. On the contrary, the third panel in figure 4
shows that the assumption of a fixed transition temperature gives
a sharp separation in the fundamentalised periods, not observed in
M3.

This evidence is supported by Kolmogorov-Smirnov test for the
distribution of c-type periods, as listed in Columns 2 and 3 of
the Table 2. The hysteresis case, out of the 9 random extractions,
reaches a maximum probability of 93.3\% against 61.3\% for a fixed
transition. The issue is, however, open to other possible
assumptions, and we present data in Fig. \ref{f:fig3} only as a
first step on the long way of a difficult investigation that will
deserve more efforts.

%%%%%%%%%%%%%%%%%%%%%%%%%%%  FIGURA 3

%%%%%%%%%%%%%%%%%%%%%%%%%%%

\section{The problem of Red HB}

The main goal of this investigation has been reached by
showing  that there is not an intrinsic and unavoidable
incompatibility between the M3 period distribution and canonical
evolutionary tracks. However, one may  go deeper, discussing
theoretical predictions in connection with the observed HB
distribution. Synthetic models in the previuos sections have been constructed by
directly relying on the observed number of variables (V) and blue
(B) HB stars. On the contrary, the number of red (R) HB stars is
a computational result, linked by evolution to the V-number.

The last column in Table 1 discloses that present
computations give in all cases a number of red HB stars larger
than the number of RR Lyrae variables. On the contrary, Catelan
(2004) quotes unpublished data for which the  530 HB stars in M3
should be distributed  according to the ratios B:V:R =
0.39:0.40:0.21. These ratios, once confirmed, raise the
contradictory evidence for which evolutionary models produce an
almost perfect fitting of the period distribution, but do not
account for the observed number of 111 red HB stars.

In this context, a firm evaluation of the HB distribution appears
of great relevance to assess the adequacy of the synthetic
procedure. However, one has to notice that on the theoretical side
the V:R ratio can be modulated in several ways, by modulating the
shape of the mass distribution and/or the temperature interval
covered by the evolutionary tracks which, in turn, for red stars
depends on several assumptions, in particular the $^{12}C(\alpha,
\gamma)^{16}O$ nuclear cross section and the free parameter
$\alpha$ governing the efficiency of convection in stellar
envelopes.

As an example, one can modify the adopted  gaussian distribution
by cutting away the  tail of  most massive stars, decreasing in
this way the number of red  stars, leaving substantially unchanged
the RR Lyrae distribution. The number of red HB stars predicted by
assuming a truncated gaussian,  i.e., neglecting all the HB masses
larger than the mean mass,  are listed in the last column of Table
3. It turns out that, over the 9 random extractions  the number of
red HB stars reaches a minimum value of 138 stars, against the 111
given by Catelan, keeping in that  case a robust 82\% of KS
probability. The semigaussian distribution
is of course an "ad hoc" assumption, but here we
are just exploring wether "ad hoc" assumptions can reconcile
canonical models  with observations. In this context, one may note
that the largely adopted gaussian mass distribution appears as an
useful and reasonable assumption. However, firm constraints on the plausibility of such a distribution
should await for new insights on the still unknown mechanism driving
the mass loss.

An overabundance of red stars could be taken also as an evidence for
the occurrence of shorter but still canonical evolutionary tracks,
as produced by passing from the adopted cross section for the $^{12}C(\alpha,\gamma)^{16}O$ reactions (Caughlan \&
Fowler 1985) to the revised values presented by Kunz et al. (2002; see also Dorman 1992).
In principle, the same effect could be produced by an increase of
the mixing length parameter $\alpha$ (see, e.g., Brocato et al..
1999). However, in our models the mixing length has been
calibrated on the temperature of the RGB branch, and a difference
of mixing length between RGB and HB stars can be barely supported.
Moreover and even  more interestingly,  we also found that a
decrease of the metallicity below Z=0.001, as  a consequence
of the recently suggested revised solar metallicity (Z$_{\odot}$ =
0.0122:  Asplund et al. 2004)  would solve the problem.

In this context, one has first to notice that such a new
value for the global solar metallicity should be not simply scaled
according to the available values of [Fe/H] for M3 in absence of
suitable 3D non-LTE models for globular clusters 
metal poor stars. However, as a first
order approximation one may take [Fe/H]=-1.5 from Kraft \& Ivans
(2003) thus deriving Z(M3) $\sim$ 0.0006 where an enhancement of
$\alpha$ elements by [$\alpha$/Fe]= 0.3 has been taken into
account. As a matter of fact, one finds that the nine
synthetic models with Z=0.0006 and a truncated gaussian predict
a mean number of red stars as given by 133,  with smaller KS probabilities, but still reaching in the
best case a number of red stars as low as 98, with a KS probability still of
the order of 30\%.

%==============================Table 2================

\begin{table}
\caption{Values of KS test for the 9 runs considering the
distribution of only RRc variables, assuming the efficiency of an
hysteresis mechanism (d) or a fixed transition temperature (e)(see
text). }

\begin{center}
\begin{tabular}{r r r }
\hline
$ Run $ & $ KS (d) $ & $ KS (e) $\\
\hline

1 &  93.35 & 61.29  \\
2 &  7.69 & 19.37  \\
3 &  18.98 &  8.44 \\
4 &  91.57 & 36.70 \\
5 &  81.22 & 39.61 \\
6 &  89.20 & 54.91 \\
7  & 43.99 & 12.90  \\
8 &  7.67 & 14.08  \\
9 &  19.05 &  6.16 \\

\hline \hline
\end{tabular}
\end{center}
\end{table}

%============================Table 3=================================
\begin{table}
\caption{Values of KS test for 9 runs with the truncated gaussian
assumption. The last column gives the number of red HB stars for
the various cases .}

\begin{center}
\begin{tabular}{r r r}
\hline
$ Run $ & $ KS (f) $ & $ N(red)$ \\
\hline

1 & 61.07   & 165 \\
2 & 37.67    & 169\\
3 & 13.24   & 149\\
4 & 4.75   & 164\\
5 & 31.25   & 204\\
6 & 61.07   & 180\\
7  & 10.40   & 159 \\
8 & 81.99  & 138 \\
9 & 37.67   & 145 \\

\hline
\hline
\end{tabular}
\end{center}
\end{table}

We are not in the position of discriminating among the various
possibilities, since both the cluster metallicity scale, the
$^{12}C(\alpha, \gamma)^{16}O$ cross section and, perhaps, the
cluster HB type are far from being firmly established. However,
the above discussion shows that the V:R ratio
does not appear to be a
firm prediction of the theory, and different paths can be
followed  to reconcile this number with observations by relying on
canonical evolutionary models.

\section{Discussion and conclusions}

As already mentioned, the canonical scenario
leaves some degree of freedom for the computed evolutionary
tracks, as a consequence of uncertainties in various physical
ingredients. On the contrary, it appears difficult to move HB
models beyond their canonical paths. The growth of the convective
cores and the efficiency of semiconvection appear the major
assumptions which could be debated, at least in principle. As well
known (see, e.g., Demarque \& Sweigart 1976) both these mechanisms
act in the sense of increasing the amount of mixing in the central
region and, in turn, the width in temperature of the loop
experienced by HB stars during their central He burning phase.
Thus a decreased amount of central convection \and or of
semiconvection appears just as an additional, alternative  way to
decrease to range of effective temperatures covered by HB
evolutionary tracks.

However, one cannot safely decrease the canonical efficiency of
these mechanisms without running against severe observational
constraints. A decrease in the efficiency of mixing
will cause indeed a decrease in
the HB lifetime with a corresponding increase of the time spent on
the Asymptotic Giant Branch. We find that  in the extreme case of
no mixing  the number ratio N$_{AGB}/N_{HB}$ of stars in the two
quoted evolutionary phase will raise towards untenable values, up
to N$_{AGB}/N_{HB} \sim$ 0.26, whereas the canonical value
N$_{AGB}/N_{HB} \sim$ 0.15 has been already proved to be in
excellent agreement with observations (Cassisi, Salaris \& Irwin
2003).

On the other hand  the  supposed occurrence of  mass loss during
the HB phase (Wilson \& Bowen 1984) can hardly be of help, since
in the case of M3 (Z$\sim$0.001) this will further increase the
range of effective temperatures covered by evolving HB models and,
in turn, the range of RR Lyrae periods.
The problem cannot be either solved by
the conjectured efficiency of an "evolutionary trapping" (Koopmann
et al. 1994) at the transition line between RRc and RRab, as
discussed by Catelan, since it would produce a peaked
distribution for both ab and c-type pulsators, not observed in M3.

Hence the supposed failure of canonical models would affect the
basis of an evolutionary theory which has already proved to finely
account for many observational features of stars evolving in our
Galaxy as in nearby ones. In this paper we have shown that
canonical HB models might account for the observed
distribution of fundamentalised period of RR Lyrae in the globular
cluster M3. We assumed a bimodal mass distribution,
with a sharply peaked mass mode just to the red side of the
instability strip which closely resembles the one discussed by
Catelan (2004) in the framework of his "trimodal" scenario.

Such a bimodal mass distribution is clearly different from the
semi-empirical one  derived by Rood \& Crocker (1989) on the basis
of the color distribution of the cluster stars, which is to our
knowledge the only one appeared in the literature. Further studies
to derive the mass distribution of the HB stars in M3 by using
more recent data could provide an important test of the bimodal
mass distribution hypothesis.

One may finally notice that the peaked period distribution
discussed in this paper for the M3 case is not a common feature of
RR periods in galactic globular clusters. We plan to discuss this
point in a following paper, which will be devoted to a synthetic
approach to RR Lyrae period distributions in both Oosterhoff I and
Oosterhoff II RR Lyrae rich clusters. Here we only notice that
inspection of Fig.1 in Paper I (Castellani, Caputo \& Castellani
2003)  reveals that  several Oosterhoff I globular clusters
show a much flatter period distributions. This is in particular
the case for the two RR Lyrae rich clusters M5 and M62.  This
occurrence requires much less severe constraints on the mass
distribution of HB stars, supporting perhaps the Catelan (2004)
suggestion for which M3 might be "a pathological case that cannot
be considered representative of the OoI class".

\begin{acknowledgements}
This research has made use of NASA's Astrophysics Data System
Abstract Service and SIMBAD database operated at CDS, Strasbourg,
France. This work was partially supported by MURST (PRIN2002,
PRIN2003). We acknowledge Giuseppe Bono for a critical reading of
the manuscript. We thank our anonymous referee for
his/her quite detailed comments and suggestions that stimulated a
relevant  widening of the original investigation.

\end{acknowledgements}

\end{document}